\documentstyle[12pt,agums]{article}
\lefthead{\'ALVAREZ ET AL.}

\righthead{Forecasting Alboran Sea with genetic algorithms}

\received{October 28, 1999}
\revised{}
\accepted{}

\paperid{}

\cpright{AGU}{1996}

\authoraddr{Alberto \'Alvarez,
SACLANT Undersea Research Centre, 19138 San Bartolomeo, La Spezia,
Italy. (e-mail: alvarez@internal.saclantc.nato.int)}

\authoraddr{Crist\'obal L\'opez, Margalida Riera,
Emilio Hern\'andez-Garc\'{\i}a and Joaqu\'{\i}n Tintor\'e,
Instituto Mediterr\'{a}neo de Estudios Avanzados, CSIC-Universitat
de les Illes Balears, E-07071 Palma Mallorca, Spain. (e-mail:
clopez@imedea.uib.es; dfsmrj4@clust.uib.es; emilio@imedea.uib.es;
dfsjts0@clust.uib.es)}

\slugcomment{To appear in the {\it Geophysical Research
Letters}, 1999.}

\begin{document}

\title{Forecasting the SST space-time variability of the
Alboran Sea with genetic algorithms}

\author{Alberto \'Alvarez}
\affil{SACLANT Undersea Research Centre,
 La Spezia, Italy}

\author{Crist\'obal L\'opez, Margalida Riera,
Emilio Hern\'andez-Garc\'{\i}a and Joaqu\'{\i}n Tintor\'e}
\affil{Instituto Mediterr\'{a}neo de Estudios Avanzados,
CSIC-Universitat de les Illes Balears, E-07071 Palma de Mallorca,
Spain}

\begin{abstract}

We propose a nonlinear ocean forecasting technique based on a
combination of genetic algorithms and empirical orthogonal
function (EOF) analysis.
 The method is used to forecast the space-time
variability of the sea surface temperature (SST) in the Alboran
Sea. The genetic algorithm finds the equations that best describe
the behaviour of the different temporal amplitude functions in the
EOF decomposition and, therefore, enables global forecasting of
the future time-variability.

\end{abstract}

\begin{article}
\section{Introduction}

Traditionally, ocean forecasting is carried out integrating
forward in time the equations of motion. This approach usually
requires the derivation of the dynamical laws controlling the
ocean processes as well as the detailed knowledge of the initial
conditions. Unfortunately, this level of knowledge or the computer
power needed for the numerical simulation of the ocean is not
always accessible. In these cases, an alternative approach to
forecast ocean evolution consists in extracting dynamical
information directly from the empirical data without imposing an
explicit dynamical model. The extracted information about the past
of the system is then used to predict its future evolution. The
classical techniques in this type of approach consist on modeling
the dynamics as a random process, using nondeterministic and
linear laws of motion
\cite{BaRo92}.
However, new techniques that explicitly take into account the
nonlinear nature of the time evolution are demonstrating a high
predictive power. Proposals based on genetic algorithms are
beginning to appear in different contexts \cite{Sz97}. Briefly
stated, genetic algorithms are methods to solve optimization
problems in which the optimal solution is searched through steps
inspired in the Darwinian processes of natural selection and
survival of the fittest
\cite{Ho92}. In the forecasting context, the optimization problem
to be solved is to find the empirical model best describing
observed past data. The empirical model so obtained may then be
used to forecast the future, and may reveal functional
relationships underlying the data. Recently,
\cite{Sz97} has already shown the robustness of genetic
algorithms to forecast the behavior of one-variable chaotic
dynamical systems.

The aim of our Letter is to extend the work of
\cite{Sz97} to spatially extended dynamical systems, thus permitting
application to real oceanographic data. More explicitly, we focus
on using genetic-algorithm methods to predict the space-time
variability of the sea surface temperature (SST) of the Alboran
Sea. The reasons for this particular election arise from the
circulation structure of this basin, the westernmost of the
Mediterranean sea, characterized by a wavelike front with two
anticyclonic gyres generated by the inflow of Atlantic waters into
the Mediterranean throught the Strait of Gibraltar
\cite{Tin91,Viu96}. This circulation pattern has a strong
signature in the SST field, which provides the chance to observe
its space-time variability from satellite imagery.

The Letter is organized as follows: Section II presents the
technique; Section III briefly describes the characteristics of
the satellite data employed in this study. The results obtained
from the application of the method are shown in Section IV, and
Section V concludes the work.

\section{Nonlinear forecasting of two-dimensional fields with
genetic algorithms}

The methods of \cite{Sz97} are adequate to forecast the time
evolution of one or a small set of variables. Two-dimensional SST
fields obtained by satellite imagery are too large data sets for
this technique to work directly. A method to encode the time
series of satellite images into a smaller set of numbers is thus
required. This can be accomplished by using the Empirical
Orthogonal Function (EOF) technique \cite{Ho96,EOFs}. Briefly, EOF
analysis decomposes the space-time distributed satellite data into
modes ranked by their temporal variance. As a result, a set of
spatial modes and associated temporal amplitude functions are
obtained. The spatial modes provide information of the spatial
structures while the amplitude functions describe their dynamics.
The complete state of the system (i.e., the original sequence of
satellite images) can be well approximated by simple linear
combination of the most relevant spatial modes multiplied by their
corresponding amplitude functions
\cite{Ho96,EOFs}. The problem of forecasting the dynamics of a
two-dimensional field has thus been reduced to predicting the
amplitude functions, a small set of time-series, corresponding to
the most relevant EOFs.

The works of \cite{Ta81}, \cite{Ca89}, and many others have
established the methodology for nonlinear modeling of chaotic time
series. Explicitly, Takens' theorem \cite{Ta81} establishes that
given a deterministic time series $\{{\bf x}(t_{k})\}, t_k=k\Delta
t, k=1,...,N$ there exists a smooth map $P
$ satisfying:

\begin{equation}
\label{uno}
{\bf x}(t)= P\left[ {\bf x}(t-\Delta t), {\bf x}(t-2 \Delta
t),\cdots, {\bf x}(t-m\Delta t ) \right]
\end{equation}

\noindent where $m$ is called the embedding dimension obtained from a
state-space reconstruction of the time series \cite{AbaBro93}.
Our aim is to obtain with a genetic algorithm the functions
$P(\cdot)$ in Eq.~(\ref{uno}) that best represents the amplitude
function associated to each one of the most representative EOFs,
and then use them to predict the future state of the system.
The algorithm proceeds as follows (for details see \cite{Sz97}):
First, for the $j$-amplitude function, $A_{j}(t)$, a set of
candidate equations (the population) for $P(\cdot)$ is randomly
generated. These equations (individuals) are of the form of
Eq.~(\ref{uno}) and their right hand sides are stored in the
computer as sets of character strings that contain random
sequences of the variable at previous times ($A_j(t-\Delta t),
A_j(t-2\Delta t), ..., A_j(t-m\Delta t)$), the four basic
arithmetic symbols ($+$,$-$,$\times$, and $/$), and real-number
constants. A criterion that measures how well the equation strings
perform on a training set of the data is its fitness to the data,
defined as the sum of the squared differences between data and
forecast from the equation string. The strongest individuals
(equation strings with best fits) are then selected to exchange
parts of the character strings between them (reproduction and
crossover) while the individuals less fitted to the data are
discarded. Finally, a small percentage of the equation strings'
most basic elements, single operators and variables, are mutated
at random. The process is repeated a large number of times to
improve the fitness of the evolving population. More details of
the algorithm are given in the Appendix.

In order to minimize the effects of the stochastic components
introduced into the amplitude functions by the measurement and
environmental noise, and by neglecting the EOFs of small variance
in the reconstruction processes, a noise-reduction method based on
Singular Spectral Analysis (SSA) or data adaptive approach
\cite{Pe91}, to be described below, is first applied to the
noisest amplitude functions.

\section{Data}

In the present study we have considered a series of 68 monthly
averaged SST images of the Alboran Sea, ranging from March-1993 to
October-1998. Each monthly image is based on the daily maximum
images using the average for every single pixel's position. The
monthly composition normally consits of approximately 160 AVHRR
passes. Several tests ensure that SST values are derived only for
cloudfree water surfaces. All pixels flagged as cloud are excluded
from all further processing. The data set is an AVHRR MCSST
product from DLR.

\section{Results}

Figure~1a, b, c and d show the mean, 1st, 2nd and 3rd spatial
modes respectively obtained from the EOF analysis, while the solid
lines in Figure 2 represent the temporal amplitude functions
associated with each spatial mode. Basically, the 1st EOF mode
captures the variability associated with the seasonal changes in
the surface temperature of Atlantic and Mediterranean waters. The
2nd spatial mode appears to be associated with variability in the
intensity of the two gyres. Finally, the 3rd mode essentially
describes the spatial variability related to the Almeria-Oran
Front. These three modes account for $98.64 \%$ of the total
variance of the data. A Complex EOF decomposition of satellite
altimetry data in the same area was performed by \cite{VaFont96}.

The amplitude functions of the 2nd and 3rd EOFs show a time
dependence much more complex than the simple seasonal variation
displayed by the 1st one. This could be an indication either of
complex deterministic evolution or of contamination from random
noise. To disentangle both components, the signals were filtered
using the SSA method: The filtered time series obtained from the
amplitude functions of the 2nd and 3rd EOFs (red dashed lines in
Figs. 2b and c) were built considering the first eight and five
SSA eigenvalues (that account for $70
\%$ and $65 \%$ of each amplitude variance) in the respective original
time series. The criterion to identify this amount of
deterministic variability in each signal was based on a nonlinear
prediction approach
\cite{Tso91}. Essentially, the signal to be filtered is rebuilt
using only a certain number of eigenvalues obtained from the SSA
decomposition. Then, the genetic algorithm is employed to find the
equation that best fits the data in one part of the dataset, the
training set, ranging from March-1993 to June-1998. The
predictability skill of the solution equation is then validated
with data ranging from July-1998 to October-1998, the validation
set, previously unknown for the algorithm. If the forecast
performance of the solution equation is high in the validation set
(more than $80 \%$ of agreement between data and forecast) the
rebuilt signal is considered to be mainly deterministic. A new
time series is then rebuilt from the original one considering a
larger number of eigenvalues and the previous process is
repeated. The procedure is stopped when the inclusion of new
eigenvalues deteriorates the forecasting skills, since then it
can be argued that the variability represented by the new
eigenvalues has a strong noisy component. The final filtered
signal is thus rebuilt with the maximum number of eigenvalues
that provide a good forecast skill in the validation set.

The resulting empirical equations obtained from the iteration of
the genetic algorithm for the three temporal amplitude functions
are written in the Appendix. Figures~2a, b and c show the results
of applying the solution equations. The blue dash-dotted line
shows the results of applying the solution equation in the
training set: all the points in the line are one-month-ahead
predictions, i.e. they are obtained from the equations in the
Appendix and the observed values of the (filtered) temporal
amplitude at $m$ previous months. Blue circles are the
one-month-ahead predicted values in the validation set, i.e., the
time interval for which measurements were used in the filtering
process but not in the final genetic forecasting.

In order to discriminate if the excellent agreement between data
and predictions in the validation set comes from artificial
dependencies in the data introduced by the filtering procedure or
from an intrinsic dynamical behavior well captured by the
evolutionary algorithm, the solution equations are tested in a 3rd
set of data called the forecasting set (from November-1998 to
January-1999) that has not been used in the filtering process. The
crosses are one-month-ahead forecasts in this set of data. The
agreement in all cases is excellent, thus indicating that the
genetic algorithm has been able to capture the main time
variability of each EOF. It is remarkable that this has been
achieved without the use of any explicit knowledge of the ocean
dynamics, and using data just from the upper layer of the sea.

It remains, to close the procedure, to obtain the total forecasted
SST spatial field. This is accomplished by adding the three EOFs
multiplied by their predicted amplitudes. This has been carried
out for the forecasting set. Figures~3a and b show the monthly
averaged SST field for November-1998 and the corresponding
one-month-ahead forecast. The result correctly reproduces the main
SST structure of the gyres in particular and the Alboran Sea in
general. The technique slightly overestimates the SST of the two
gyres. Since the agreement with the filtered time-series was
rather good, this discrepancy should correspond to the part of the
observation that has been identified as stochastic by the
algorithm. The results obtained for December-1998 are shown in
Figures~3c and d. In this case, the forecasted field still keeps a
slight signature of the two gyres, a feature that is not found in
the real data, although the presence of warm Atlantic waters in
the gyre areas is well reproduced. Finally, Figures~3e and f
describe the results obtained for January-1999. The real as well
as the forecasted fields show a general cooling of the basin with
the disappearance of the Alboran gyres.

\section{Conclusion}

We have proposed a new technique that allows prediction of ocean
features using satellite imagery. We first compute the dominant
spatial EOF modes from a time series of satellite data, and next
we forecast their time evolution using genetic algorithms. The
technique has been applied to the one-month-ahead prediction of
the SST field of the Alboran sea, and has demonstrated a good
performance. We expect the method to perform well in any other
situations in which ocean structures are sufficiently permanent
for the EOF method to provide large data compression, and in which
the dominant EOFs contain a strong deterministic-evolution
component as compared to the stochastic one. The method can be
applied to any field observable from satellite (SST, dynamic
height, surface ocean colour), and the information obtained could
be useful for operational needs such as fisheries, naval
operations and even for assimilation into numerical models.

\appendix

\section{Appendix A: Analytical expressions for EOF amplitude functions}

The values of the  parameter $m$ are $m=6, 12$ and $12$ for the
1st, 2nd and 3rd EOF respectively and the maximum number of
symbols allowed for each tentative equation is 20. Each generation
consists of a population of 120 randomly generated equations.
After $10000$ generations we obtain for the 1st, 2nd and 3rd
amplitude functions the following expressions:

\begin{equation}
\label{dos}
A_{1}(t) =0.33 \left(2~A_{1}(t-1)-\left(A_{1}(t-3)+A_{1}(t-6)+
\left({A_{1}(t-1) \over \left(-3.78-{\left(A_{1}(t-1)-9.3 \right)
\over A_{1}(t-2) }\right)} \right) \right) \right).
\end{equation}

\begin{eqnarray}
\label{tres}
&&A_{2}(t) =A_{2}(t-1) -A_{2}(t-2)
\nonumber \\
&&-0.134 \left(A_{2}(t-4)-\left(A_{2}(t-5)-A_{2}(t-12)
-3.45\left(A_{2}(t-5)+A_{2}(t-8)\right) \right) \right) .
\end{eqnarray}

\begin{equation}
\label{cuatro}
A_{3}(t) ={0.4 A_{3}(t-12)} -0.4 - { 0.59
\left(2.5-A_{3}(t-3)+A_{3}(t-9)-A_{3}(t-1)\right)}.
\end{equation}

\acknowledgments
 Financial support from CICYT (AMB95-0901-C02-01-CP and
 MAR98-0840), DGICYT (PB94-1167), and from the MAST program MATTER
 MAS3-CT96-0051 (EU) is greatly acknowledged.


{}
\end{article}

\newpage

\begin{figure}
submitir
        \caption{a) Mean SST of the Alboran Sea; b), c) and d)
shows the 1st, 2nd and 3rd EOF respectively.}
\end{figure}

\begin{figure}
        \caption{a) Amplitude function corresponding to the 1st EOF (solid
black line); the blue dash-dotted line shows the results of
applying the solution equation in the training set while
the crosses are one-month-ahead forecasts in the forecasting data set.
b) The solid black line represents the observed amplitude of the 2nd EOF.
The red dashed line represents the SSA-filtered mode amplitude.
The dash-dotted line represents the fitting of the solution
equation to the SSA-filtered mode amplitude in the training set.
Circles and crosses represent one-month-ahead forecasts in the
validation and forecasting data sets, respectively; c) same as b)
but for the 3rd EOF.}

\end{figure}

\begin{figure}
        \caption{a) Monthly mean SST of the Alboran Sea
corresponding to November-1998 and b) forecast obtained for
November-1998 one month in advance; c) Monthly mean SST of the
Alboran Sea corresponding to December-1998 and d) forecast
obtained for December-1998; e) Monthly mean SST of the Alboran Sea
corresponding to January-1999 and f) forecast obtained for this
month.}

\end{figure}

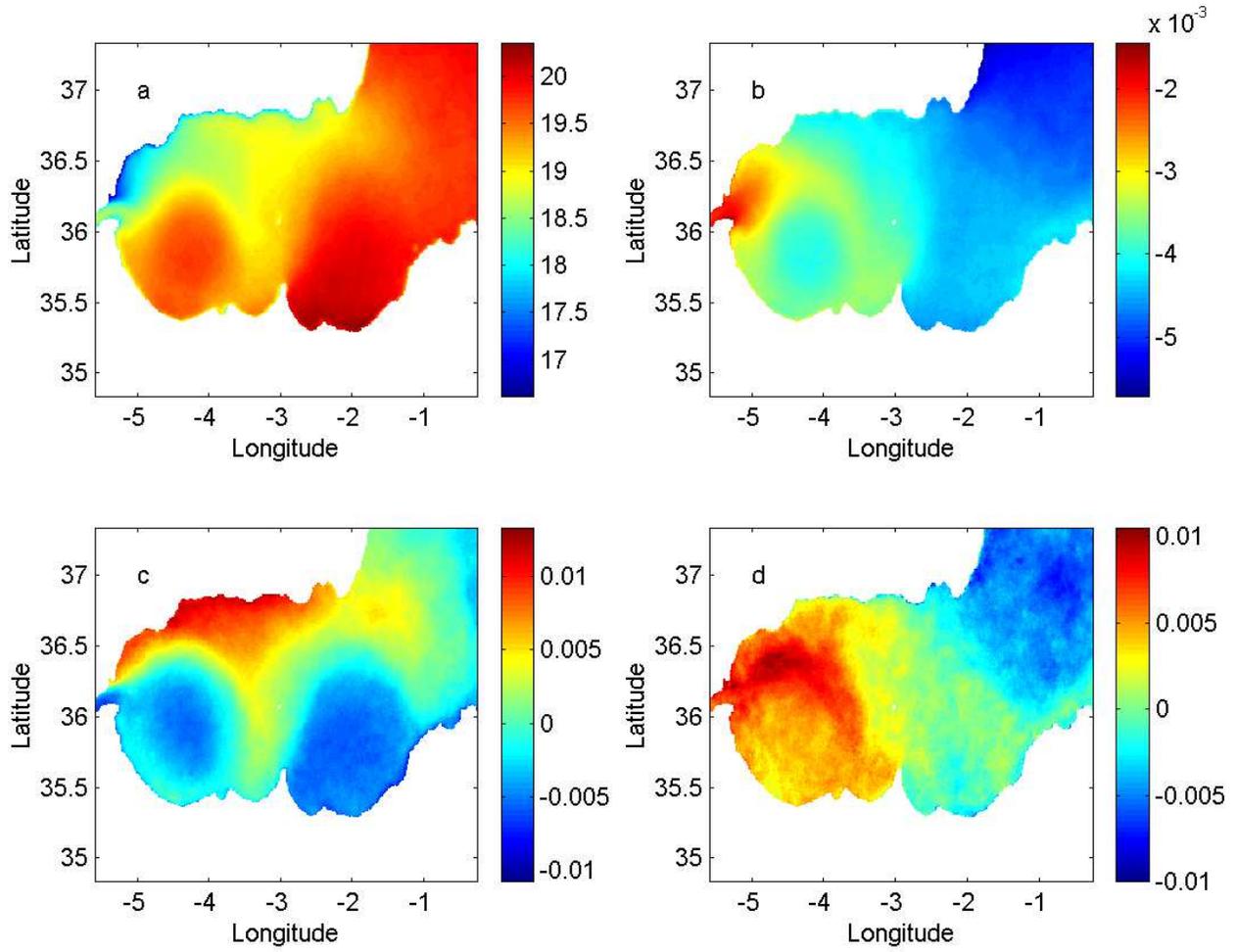
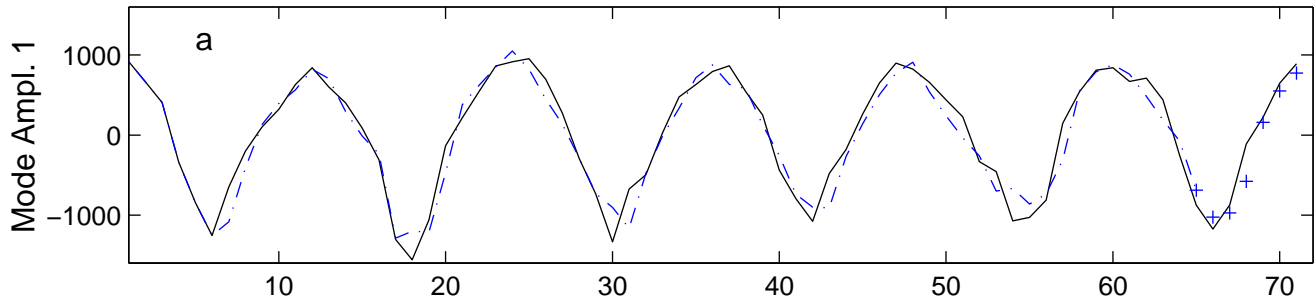
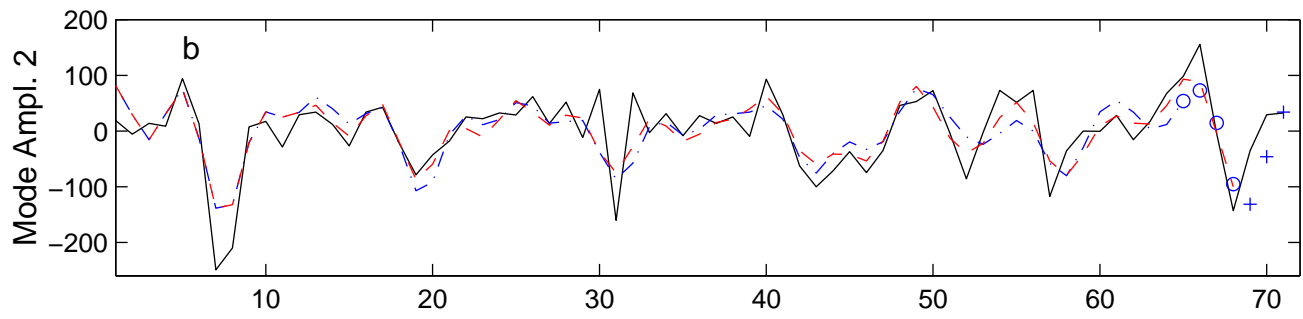
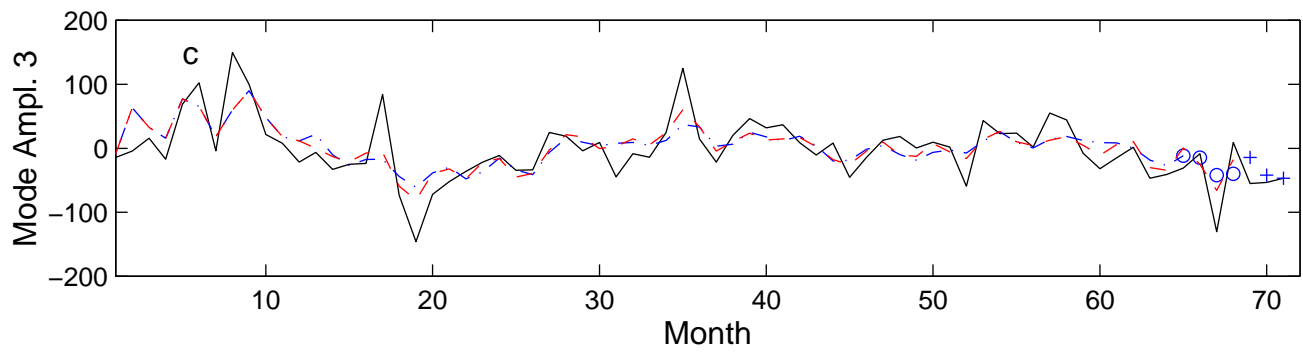
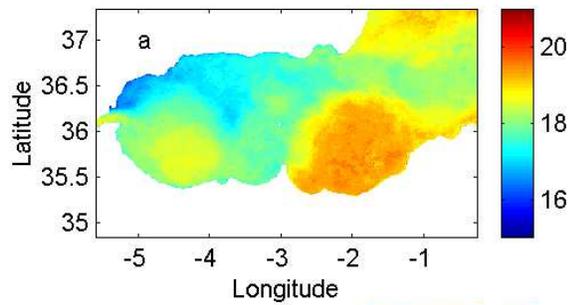
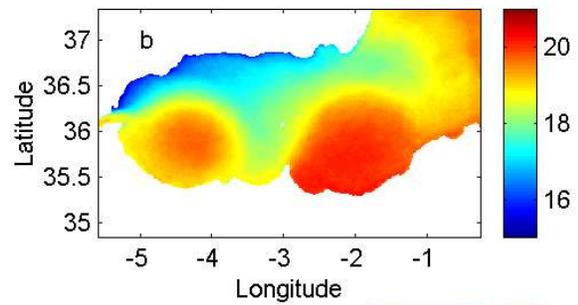
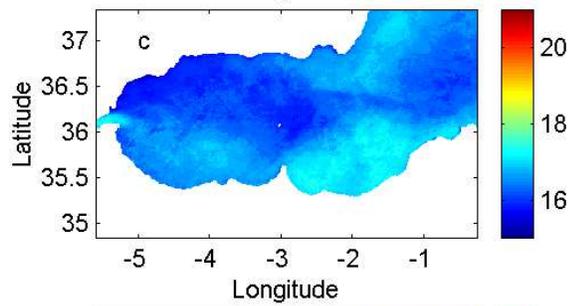
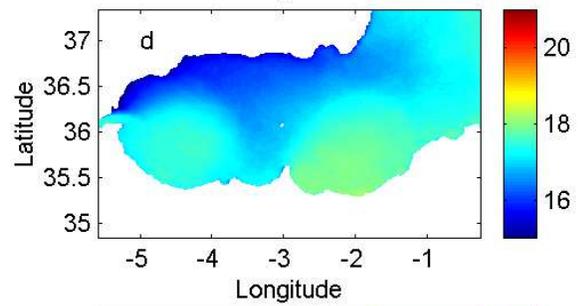
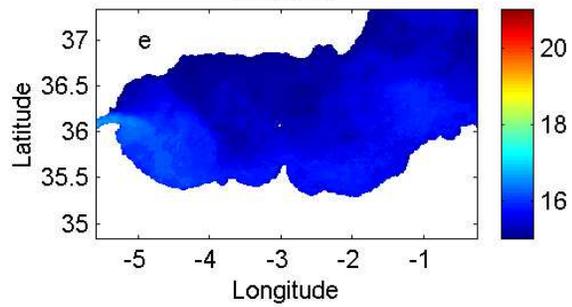
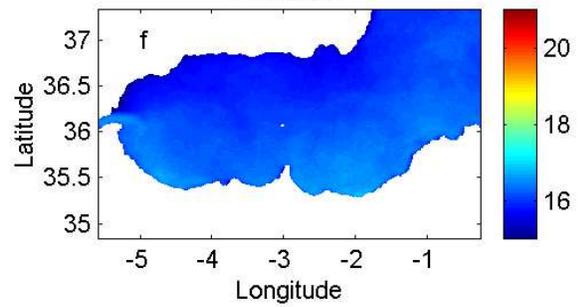
\end{document}